\documentstyle[preprint,prb,aps]{revtex}
\def\dps{\displaystyle}
\def\ivn{i\nu_n}
\def\intuu{\int_{-\infty}^\infty}
\def\PNCAU{$\mbox{{\em PNCA}}_\infty$}
\draft
\begin{document}
\title{An Enhanced  Perturbational Study on Spectral Properties of
the Anderson Model}
\author{F B Anders \\
Institut f\"ur Festk\"orperphysik, \\
Technische Hochschule D-64289 Darmstadt,
Germany}
\date{February 3, 1995}
\maketitle
%%
%% Abstract
%%
\begin{abstract}
The infinite-$U$ single impurity Anderson model for rare earth alloys is
examined with a new set of self-consistent   coupled integral
equations, which can be embedded in  the large $N$ expansion scheme ($N$ is
the local spin degeneracy).
The finite temperature impurity density of states (DOS) and the
spin-fluctuation spectra are calculated exactly up to the order
$O(1/N^2)$.
The presented conserving approximation goes well beyond the
$1/N$-approximation ({\em NCA})  and maintains local Fermi-liquid properties
down to very low temperatures.
The position of the low lying Abrikosov-Suhl resonance (ASR) in the impurity
DOS  is in accordance with Friedel's sum rule. For $N=2$ its
shift toward the chemical potential, compared to the {\em NCA}, can be traced
back  to the influence of the  vertex corrections.
The width and height of the ASR is governed by the universal low
temperature energy scale $T_K$.
 Temperature and degeneracy $N$-dependence of the static magnetic
susceptibility is found in excellent agreement with the Bethe-Ansatz
results.   Threshold exponents of the local propagators are discussed.
 Resonant level regime ($N=1$) and intermediate valence regime ($|\epsilon_f|
<\Delta$) of the model are thoroughly investigated as a critical test
of the quality  of the approximation.
Some applications to the Anderson lattice model are pointed out.
\end{abstract}
\pacs{PACS numbers: 71.28, 75.20H}

\section{Introduction}

Recently a major improvement has been reported on  the perturbational
approach to the single impurity Anderson-model ({\em SIAM}) with
$N$-fold degenerate local state  in the limit of infinitely large
Coulomb repulsion $U$\cite{AndersGre94}.
The aim of this paper is to present in detail the results obtained so
far with  this so-called Post-{\em NCA} theory (\PNCAU). This approximation
contains a resumation of infinite numbers of skeleton diagrams and is
exact up to order $O(1/N^2)$. The skeleton diagrams included
physically describe complicated multi-excitation processes.

The paper is
organized as follows: in section 2, the large-$N$ expansion is used
to derive the self-consistent coupled integral equation for the
self-energies and the vertex-function. The physical processes
included will be explained. In section 3  numerically obtained
solutions for the local propagators will be presented and their
threshold exponents related to results furnished by other methods.
A detailed analysis of the one-particle spectra will be the topic of
section 4 including a discussion of the Fermi-liquid properties and
the impact of the vertex-corrections. In section 5, we will develop
a theory of the magnetic vertex correction exact up to $O(1/N^3)$ and
compare the calculated static magnetic susceptibility with the
Bethe-Ansatz results.  Spin-fluctuation spectra will also be
covered. The two critical cases, the
intermediate-valance and the resonant level case, will be thoroughly
examined in section 6.  In section 7, a summary and an outlook on
further applications will be given.

\section{Theory}

In direct perturbation theory  a well established
set of diagrammatical rules has been established to evaluate self-energy
contributions to the local propagator $P_M(z)$
\cite{KeiterMor84,Bickers}. In Fig. ~\ref{fig-sig0}
the lowest
order contributions up to $V^8$ to the
self-energy $\Sigma_0(z)$ of  propagator $P_0(z)$ of the un-occupied
state are shown. Contributions (b) and (c) are one-particle
irreducible but contain self-energy insertions and are  already
 included in the skeleton diagram (a). In order $V^6$  there
exist  only {\em one} skeleton diagram, whereas there are  exactly
{\em two} in order $V^8$. To include these higher order
corrections, a well-established concept of vertex functions is used. In
our case, a vertex function  is  defined by cutting the electron and
the local propagator line at  the uppermost vertex in each skeleton
diagram   and by summing up all contributions of the remaining
lower parts of the skeletons. The concept of vertex-function has already been
successfully applied in an {\em NCA}-theory
(Non-Crossing-Approximation) for the finite-$U$ Anderson
model\cite{PruschkeGre89}. The resulting dimension-less factor
$\Delta_m^{I}(x,y)$   renormalizes the
bare hybridisation vertex $V$ of an absorbed band electron in an
energy dependent way; $x$ is the ingoing and $y$ the outgoing energy
of the local propagator.  The corresponding vertex-function
$\Delta_m^{O}(x,y)$ for an emitted band electron  is obtained from
the vertex function of an absorbed electron $\Delta_m^{I}(x,y)$ by
symmetry: $\Delta_m^{O}(x,y) = \Delta_m^{I}(y,x)$.
Therefore we only  calculate
$\Delta_m(x,y) \equiv \Delta_m^{I}(x,y)$.  In a formal way all
skeleton diagrams are summed up  by using the exact vertex-function in:
\begin{eqnarray}\label{equ-sig-0}
\displaystyle \Sigma_0(z)  & = &\displaystyle  V^2\sum_{km}
f(\epsilon_k) \Delta_m(z,z+\epsilon_k)P_m(z+\epsilon_k)
\\[5pt]
\label{equ-sig-1}
\displaystyle \Sigma_m(z)  & = &\displaystyle  V^2\sum_{k} f(-\epsilon_k)
\Delta_m(z-\epsilon_k,z)P_m(z-\epsilon_k)\;\; .
\end{eqnarray}
Approximations are made by the selection of  contributions to
$\Delta_m$. The {\em NCA}, for example, is recovered by setting
$\Delta_m(x,y) =1$.

In order to derive the integral equation for the vertex function
exactly up to $O(1/N^2)$  the lower parts of the skeleton diagrams
(d) and (e)  have to be included in  addition to skeleton diagram
(a). Skeleton (f) can be neglected,
since it is of the order $O(1/N^3)$. The corresponding generating
functional is shown diagrammatically in Fig.~\ref{fig-gen-func}.
An infinite number of skeletons are resummed by replacing  the bare
hybridisation vertex  by the renormalized one $V\cdot\Delta_m(x,y)$ in each
diagram in order to include as many higher order terms as possible.
This yields  a self-consistent equation for the vertex
function displayed  in
Fig.~\ref{fig-vertex-voll}\cite{AndersGre94}. In the case of zero
magnetic field all $P_m$ are equal, and the vertex function is
reduced to
\begin{equation}
\label{equ-delta-aux}
\begin{array}{l}
\Delta(x,y) \equiv
\displaystyle \Delta_m(x,y) \; = \\
\begin{array}{cr}
1\; - & \displaystyle |V|^4\intuu\intuu dudv \rho(u)\rho(v)f(u)f(-v)
 G(x,x+u,x+u-v)\\
& \cdot \Delta(x+u-v,y+u-v) H(y,y-v,y+u-v) \; \; .
\end{array}
\end{array}
\end{equation}
Here we have chosen $\rho(\epsilon)$ to be the  density-of-states
(DOS) of the conduction electrons and the
auxiliary functions $G$, $H$ and $K$
\begin{eqnarray}\label{equ-delta-aux-g}
  G(x,x+u,x+u-v)&   = &  \begin{array}[t]{l}
  \displaystyle\Delta(x,x+u)P_m(x+u) \\
 \Delta(x+u-v,x+u)P_0(x+u-v)
\end{array}
\\
\label{equ-delta-aux-h}
  H(y,y-u,y+u-v) & = &
\begin{array}[t]{l}
P_m(y+u-v) \\
\begin{array}[t]{r}
\cdot(\Delta(y-v,y+u-v)P_0(y-v)\Delta(y-v,y) \\
- K(y+u-v,y,y+u))
\end{array}
\end{array}
\\[20pt]
%
% begin j
%
\label{equ-delta-aux-j}
K(y+u-v,y,y+u)\; & = &
 \begin{array}[t]{r}
\displaystyle N |V|^2\intuu dl \rho(l)f(-l)
\Delta(y+u-v-l,y+u-v) \\
\cdot \displaystyle P_0(y+u-v-l)\Delta(y+u-v-l,y+u-l)\\
\displaystyle \cdot P_{m}(y+u-l) \Delta(y-l,y+u-l)\\
\displaystyle \cdot P_0(y-l) \Delta(y-l,y) \;\; .
 \end{array}
\end{eqnarray}
 The negative sign in Eqn.(\ref{equ-delta-aux})
takes into account the odd number of crossing electron lines in the
skeleton (d) while in diagram (e)  an even number of crossing lines
can be found. The  approximation defined by Eqn. (\ref{equ-sig-0}) to
(\ref{equ-delta-aux-j})  will be called \PNCAU\  in the
following ($\infty$ stands for $U=\infty$).

In the  one-particle
Green's function $F_m(i\omega_n)$ the vertex correction also comes
into play in a natural way\cite{Grewe83}:
\begin{equation}\label{equ-gf}
F_m(i\omega_n) = \frac{1}{Z_f}\oint_{{\cal C}}\frac{dz}{2\pi
i}e^{-\beta z} \Delta_m(z,z+i\omega_n)P_0(z) P_m(z+i\omega_n)\; \; .
\end{equation}
This equation  can e.~g.~be obtained by cutting one band electron
line in the functional $\Phi[P_M]$ of Fig.~\ref{fig-gen-func}. Thereby, no
 ambiguity is left in the analytic expression.
The contour ${\cal C}$ encircles the all singularities of the kernel
in a counterclockwise fashion. Analytic continuation of the Green's
function   gives
\begin{equation}
\label{equ-schnitt-gfsiam}
\begin{array}[b]{lcl}
F_{m}(\omega+i\delta) & = &
\displaystyle\intuu dw' P_m(\omega +w')
\begin{array}[t]{l}\displaystyle(\Re\Delta_x(\omega,\omega+w')\xi_0(w') \\
\displaystyle + \Re e[P_0(w')]\xi_{\Delta_x}(\omega,\omega+w'))
\end{array}\\[20pt]
& - &
\displaystyle\intuu dw' P_0^\star(w'-\omega)
\begin{array}[t]{l}\displaystyle (\Re\Delta_y(w'-\omega,\omega)\xi_m(w')\\
\displaystyle + \Re e[P_m(w')]\xi_{\Delta_y}(w'-\omega,w'))\;\; .
\end{array}
\end{array}
\end{equation}
We have introduced the real auxiliary functions
\begin{equation}
\xi_M(\omega)  \equiv  -\frac{1}{\pi Z_f}
 e^{-\beta\omega} \Im m P_M(\omega +i\delta)
\end{equation}
and the complex auxiliary functions
\begin{equation}\label{gf-cuts}
\begin{array}{lcrcl}
\Re\Delta_x(\omega,\omega+i\omega_n) & \equiv &
\frac{1}{2}[\Delta(\omega+i\delta,\omega+i\omega_n)
& + & \Delta(\omega-i\delta,\omega+i\omega_n) ]\\
\Im\Delta_x(\omega,\omega+i\omega_n) & \equiv &
\frac{1}{2i}[\Delta(\omega+i\delta,\omega+i\omega_n)
& - & \Delta(\omega-i\delta,\omega+i\omega_n)] \\
\Re\Delta_y(\omega-i\omega_n,\omega) & \equiv &
\frac{1}{2}[\Delta(\omega-i\omega_n,\omega+i\delta)
& + & \Delta(\omega-i\omega_n,\omega-i\delta)] \\
\Im\Delta_y(\omega-i\omega_n,\omega) & \equiv &
\frac{1}{2i}[\Delta(\omega-i\omega_n,\omega+i\delta)
& - & \Delta(\omega-i\omega_n,\omega-i\delta) ]\\
\xi_{\Delta_x}(x,y) &\equiv &
-\frac{1}{\pi Z_f} e^{-\beta x}\Im \Delta_{x}(x,y) && \\
\xi_{\Delta_y}(x,y) & \equiv &
-\frac{1}{\pi Z_f} e^{-\beta y}\Im\Delta_{y}(x,y) && .
\end{array}
\end{equation}
The  functions $\xi_M(\omega)$  introduced above have a simple
physical meaning as {\em defect propagator}: integration over the whole
frequency range
yields the occupation probability $<\hat X_{MM}>$ of the local state $M$.
The complex functions $\Re (\Im) \Delta_{x(y)}$ and
$\xi_{\Delta_{x,y}}(x,y)$ enable the evaluation
of Eqn.~(\ref{equ-gf}) for real frequencies and thus serve purely
numerical purposes.

\section{Local Propagators, Threshold exponents and Specific Heat}
\label{sec-ion}

On a cluster of work-stations, we brought a numerical iteration
procedure for  the above system of integral equations to full convergence
using dynamically defined logarithmic meshes for the threefold
integration in (\ref{equ-delta-aux}). Besides the modulus $||P_M^{(i)}
- P_M^{(i+1)}||$ ($i$ labels the step of iteration), which reaches a
value of typical $10^{-16}$ at the end, the sum rules
\begin{eqnarray}
%%\label{equ-p-sum}
\displaystyle \oint_{{\cal C}} \frac{dz}{2\pi i} P_M(z) =  1 \;\; ; \;\;
\displaystyle \oint_{{\cal C}} \frac{dz}{2\pi i} \Sigma_m(z)  =
\displaystyle |V|^2\sum_{km}\begin{array}[t]{l}
\displaystyle \left(<M|\hat n_m^f|M>[1-f(\epsilon_{km})] \right. \\
\displaystyle \left. <M|(1-\hat n_m^f)|M>f(\epsilon_{km}) \right) \\
\end{array}
\label{equ-self-sum}
\end{eqnarray}
have been checked to estimate the quality of the numerical
calculations. The deviation between left and right side is in the
range of typically 1-4 \% and scales with the inverse number of mesh
points.
 For all numerical studies all
energies are measured in units of the Anderson width $\Delta=\pi
V^2{\cal N_F}$ ($V^2$ is the square of the hybridisation matrix
element and ${\cal N_F}$ the band DOS at the chemical potential). The
featureless symmetric conduction band DOS has been
chosen  to be $\rho^c(\omega) =
\frac{1}{2\Gamma(1.25)W}\exp(-(\omega/W)^4) $ to reduce band edge
effects.  The half band width is set to $W =10\Delta$. In most cases the
temperature will be measured in units of the corresponding Kondo-energy
\begin{equation}\label{equ-tk}
  T_K = W\left(\frac{\Delta}{\pi W}\right)^{\frac{1}{N}} \exp\left(-
\frac{\pi|\epsilon_f|}{N\Delta}\right)\;\; .
\end{equation}

For the investigation of the Kondo-regime $\epsilon_f = -3\Delta$
and $N=2$ is chosen, since the largest impact of the vertex function
is to be expected for small $N$. In Fig.~\ref{fig-prop-0-m} the
spectral density $\displaystyle\rho_0(\omega)\equiv \frac{1}{\pi}\Im m
P_0(\omega -i\delta)$ is displayed  in
the vicinity of the threshold for three different temperatures
$T= 0.5, 1, 2 T_K$. Note the fact that the energy scale has
been shifted by the threshold energy $E_s$.
The  transformation $Z_f = \oint\frac{dz}{2\pi i}e^{-\beta z}P_M(z)
\equiv e^{-\beta E_s}\tilde Z_f$
defining a renormalized local partition function $\tilde Z_f=
\oint\frac{dz}{2\pi i}e^{-\beta z}\tilde P_M(z)$ and a
local propagator $\tilde P_M(z) \equiv P_M(z+E_s)$,  ensures that
the numerically calculated $\tilde Z_f$  stays of the order $O(1)$
during the variation of temperature.
Also the {\em defect propagators} $\xi_M$ can be calculated very
accurately down to very low temperatures.

With decreasing  temperature
the threshold behaviour of the  propagator $P_0$ in \PNCAU\  exhibits a weaker
increase than the corresponding {\em NCA} propagator.  We regard this as a
strong hint toward a reduced threshold exponent $\alpha_0$ compared
with the {\em NCA} due to the influence of the vertex corrections. On the
other hand, the inset indicates quite clearly an  excellent
agreement on the high energy part of the spectrum in both
approximations as to be expected.
The spectral density $\displaystyle\rho_m(\omega)$, shown in
Fig.~\ref{fig-prop-0-m}b
for the same set of model parameters and
temperatures, reveals an additional difference between the two
approximations. $P_0^{PNCA}$  and $P_m^{PNCA}$ develop their common
threshold energy $E_s$ already at finite temperature while in {\em NCA} an
identical $E_s$ for both propagators is found only at $T=0$. This
gives rise to a temperature dependent position of the ASR in
the one-particle spectra in {\em NCA}, as we will see in the next section.

To obtain a first estimate for  threshold exponents we fit
the ionic spectrum in the low frequency range $0< \omega <T_K$ to a
trial function $h_M(\omega) = a(T)\cdot \omega^{-\alpha_M(T)}$ at
different temperatures. This procedure has been checked for
the {\em NCA} and provides the {\em exactly} known {\em NCA} exponents
$\alpha_0
=\frac{N}{N+1}$ and $\alpha_m=\frac{1}{N+1}$ within an accuracy of
$5\%$. While no temperature dependence is found for the exponent
$\alpha_0 = 0.44\pm 0.02$, Fig.~\ref{fig-schwell-a0}a,
the different values of
$\alpha_m(T)$ have been used to to extrapolate $\alpha_m(0) = 0.27\pm
0.01$ for the present parameters $N=2, \epsilon_f = -3\Delta$ giving
$n_f = 0.87$, Fig.~\ref{fig-schwell-a0}b.

There is still no agreement in the literature about the
exact exponents for $N\ge 2$. The proposal of Menge and
M\"uller-Hartmann\cite{MengeMue88}
\begin{equation}\label{equ-a-menge}
  \alpha_0 = \frac{n_f^2}{N} \hspace{20mm} \alpha_m = \frac{n_f}{N}(2-n_f)
\end{equation}
indicates a dependence of the exponents not only of the
degeneracy $N$ but also on the occupation number $n_f$ in all
regimes. This proposal has been recently backed by a  analysis of
pseudo-boson and fermion propagators provided by the numerical
renormalisation group \cite{CostiSchmiKroWoe94}.
On the other side, an analysis of parquet-equations
for the Kondo limit of the model ($n_f =1$)  claims
that the exponents
\begin{equation}\label{equ-a-grune}
  \alpha_0 = \frac{N-\frac{2}{N^2}}{N+1-\frac{2}{N^2}} \hspace{20mm}
\alpha_m = \frac{1-\frac{1}{N^2}}{N+1-\frac{2}{N^2}}
\end{equation}
should be exact in order $O(1/N^2)$ \cite{GrunebergKei91}.
We did not investigate the threshold behaviour on the
full scale of the model parameters. Therefore we cannot rule out any
of the proposals, even though the \PNCAU\  exponents in the
Kondo-regime come very close to Gruneberg's and Keiter's results.

Since we have convinced ourselves that the high energy parts of the
spectra remain unchanged, we can still use the {\em NCA} to obtain specific
heat data for a wide temperature range. On the other hand, the low
energy corrections turn out to be essential for the calculation of
correlation functions. In Figs.~\ref{fig-sh-nca-n2-tk}
 and \ref{fig-sh-nca-n2-delta0}  origin and scaling behaviour of the
two different
contributions to the specific heat are clearly demonstrated. In (a) the
effective hybridisation is tuned leading to a renormalized Anderson width
$\Delta^\ast$ and in (b) the position of the bare $f$-electron level
has been varied for $N=2$. While the first maximum in each curve of
the specific heat turns out to be an universal function of $T/T_K$,
the position of the second rather determined by the $f$-level energy
and its width by $\Delta^\ast$. Therefore, the first peak is generated by
spin excitations on the Kondo-scale $T\sim T_K$, whereas
the second contains charge excitations from the broadened $f$-level into
unoccupied band states. With increasing degeneracy the  maximum due
to spin excitations is enhanced essentially linearly with $N$ \cite{Anders95},
since their contribution is proportional to the number of channels, as
it is already know from Bethe-Ansatz calculations \cite{Rajan83}. The
charge excitations peak is rather decreased and smeared out by
broadening of the $f$-level proportional to $N\Delta$. Its position
and broadening is in good agreement with Bethe-Ansatz
\cite{okijiKawa84} and numerical renormalisation group calculations
\cite{CostiHewZla94}, but the height comes out somewhat smaller. This
could be due to absence of the doubly occupied state for
$U=\infty$. Of course, the appearance of the charge excitations  is
more of academic interest for the low temperature behaviour, but may
be important for spectroscopic studies.

\section{One-particle Spectra}\label{sec-green}

The generalized Friedel`s sum rule
\begin{equation}
\label{equ-gen-friedel}
n_f(T) = \frac{1}{\pi}\sum_m\intuu d\omega \left(\partial_\omega
f(\omega)\right) \delta_m(\omega)
+
\underbrace{\frac{1}{2\pi i}\sum_m \oint_{{\cal C}} dz f(z)
F_m(z)\partial_z\Sigma_m(z)}_{\equiv \kappa(T)}
\end{equation}
relates the local occupation number $n_f$ to the generalized phase shifts
$\delta_m(\omega) \equiv - \Im m
\ln\left[-F_m^{-1}(i\delta)\right]$. $\kappa$ essentially measures the
asymmetry of the $f$-electron spectrum and can be interpreted as the
negative change of the number of band electrons in the presence of the
impurity\cite{AndersGreLor91}. In the limit $T=0$,
Eqn.(\ref{equ-gen-friedel}) is used in combination with local
Fermi-liquid relations to derive the density of states  rule:
\begin{equation}
\label{equ-dos-rule}
\rho^{(f)}(0,T=0) =  \frac{1}{\pi \Delta}
\sin^2\left(\frac{\pi(n_f-\kappa)}{N} \right)\;\; .
\end{equation}
This formula predicts a scattering resonance with a maximum height of
$\frac{1}{\pi\Delta}$ near the chemical potential ($(n_f -\kappa)\approx 1$ )
for $N=2$  and $T=0$, the ASR,
which moves away form $\mu$
increasing degeneracies $N$. Fig.~\ref{fig-asr-n2}
shows the
one-particle spectra of \PNCAU\  and {\em NCA} for five values of temperatures.
The most significant differences between both approximations concern the
position and height  of the ASR. The position in {\em NCA} is found strongly
temperature
dependent and the height  exceeds the unitarity limit of
$1/(\pi\Delta)$ already at temperatures slightly below $T_K$. We can trace back
the temperature dependence to the mismatch of the threshold energies of
$P_0$ and $P_m$, which merge only at $T=0$. The violation of
Eqn.~(\ref{equ-dos-rule}) clearly indicates the importance of the
vertex corrections. On the other hand, the \PNCAU-ASR grows with
decreasing temperature at a stable position close to the chemical
potential. It violates only sightly the density of states sum rule
and its agreement with Friedel's sum rule (\ref{equ-gen-friedel}) is
found to be within $7\%$ as indicated in table \ref{tab-friedel-pnca}.

By focusing our attention on Fig.~\ref{fig-asr-vertex}, we gain insight
on the impact of  vertex corrections on the the ASR. The \PNCAU\
result (solid curve) is compared to the {\em NCA} result (dashed curve) for
the fixed temperature $T=0.5T_K$ and additionally to a curve where the
{\em NCA} propagators $P_M(z)$ have been used to evaluate the renormalized
hybridisation $V\cdot \Delta(x,y)$ via Eqn.(\ref{equ-delta-aux}). In
the later convergence is achieved after two steps of iteration.  The
resulting ASR, calculated after each step, has been shifted
towards the chemical potential, but stays enhanced compared to
the \PNCAU-ASR. The reduction of height of the ASR is clearly
connected to the modified threshold exponents as is also revealed by
an inspection of the iteration procedure. This stressed the importance
of vertex corrections in the local propagators $P_M(z)$.

The plots in figure \ref{fig-im-self} attract attention to the  local
Fermi-liquid properties as seen in the imaginary part of the self-energy
of the Green's function:
\begin{equation}
\label{equ-im-slf}
\Im m\Sigma_{fm}(\omega-i\delta) = \Delta + {\cal C_\omega}\omega^2 +
 {\cal C_T}T^2 \;\; .
\end{equation}
{}From the symmetric Anderson model it is know that ${\cal C_\omega} =
\Delta/T_K^2$ and  ${\cal C_T} = \Delta\frac{\pi^2}{T_K^2}$
\cite{Hewson93}. In the strongly asymmetric case ($U=\infty$) under
consideration here, only the scaling ${\cal C_\omega} = \Delta/T_K^2$
can be found. The quadratic expansion coefficient
for the temperature exhibits a rather strong dependence on the
degeneracy $N$ as been shown in Fig.~\ref{fig-im-self}, originating
from the shift of the ASR away from the chemical potential, while in
the completely symmetric case the phase shift $\delta_m$ remains
$\pi/2$ independently of $N$. On the base of our \PNCAU-results we
suggest a new analytic investigation of Fermi liquid relations for the
asymmetric case.

\section{Magnetic Susceptibility and Spin-Fluctuations}
\label{sec-mag-sus}

 Magnetic excitations are measured  by the magnetic susceptibility:
\begin{equation}\label{equ-sus-bos}
\chi(\ivn) = -\frac{1}{Z}\oint_{{\cal C}}\frac{dz}{2\pi i} e^{-\beta
z} \mbox{Tr}\left[\frac{1}{z-\hat H}\hat M\frac{1}{z+\ivn -\hat
H}\hat M \right] = \chi(-\ivn)
\;\; ,
\end{equation}
where $\ivn = \frac{2\pi i n}{\beta}$ is a bosonic
Matsubara-frequency. Focusing our attention on the operator the local
magnetization in the SIAM, $\hat M_z \equiv g\mu_B \sum_m m\cdot\hat
X_{m,m}$, we rewrite $\chi_f(\ivn)$ in analogy to  $F_m$:
\begin{equation}\label{equ-sus-vertex}
  \chi_f(\ivn) = -\frac{N\mu_j^2}{3} \frac{1}{Z}\oint_{{\cal
C}}\frac{dz}{2\pi i} e^{-\beta z} P_m(z) P_m(z+\ivn) \Gamma(z,z+\ivn)
\;\; ,
\end{equation}
defining $\mu_j^2 \equiv j(j+1)(g\mu_B)^2$. The new magnetic vertex
function $\Gamma(x,y)$ formally includes all higher order contributions arising
from diagrams with crossing band electron lines; in {\em NCA} $\Gamma(x,y)=1$.
It
turns out to be symmetric in its complex energy arguments in order to maintain
the symmetry $\chi(\ivn) = \chi(-\ivn)$. Using the standard set of
diagrammatical rules \cite{Bickers}, we obtain $\Gamma(x,y)$ exactly
up to order $O(1/N^3)$ shown in Fig. \ref{fig-mag-sus-vertex}.
$O(1/N^2)$ contributions compensate each other.
On the other hand, diagram (c) and (d) in
Fig.~\ref{fig-mag-sus-vertex} transform into each other by exchanging
the energy arguments. Therefore, we only have to insert
\begin{eqnarray}\label{equ-chi-7}
 \Gamma_c(x,y) &=  &  \dps - |V|^4\intuu\intuu dudv
\rho(u)\rho(v) f(-l)f(-u)f(v) \nonumber \\
  &  & \dps\cdot P_0(x-u)\Delta(x-u,x) \nonumber \\
 & &  \cdot P_m(x-u+v)\Delta(x-u,x-u+v) \nonumber \\
 & & \cdot\Gamma(x-u+v,y-u+v) \nonumber \\
 && \cdot P_m(y-u+v) j(y+u-v,y,y+u)
\end{eqnarray}
into the magnetic vertex function of the \PNCAU-theory
\begin{equation}
  \Gamma(x,y)  =  1 + \Gamma_c(x,y) + \Gamma_c(y,x) \;\; ,
\label{equ-chi-8}
\end{equation}
and to solve Eqn.~(\ref{equ-chi-7}) and
(\ref{equ-chi-8}) self-consistently.  Eqn.(\ref{equ-sus-vertex}) then
leads to a  local dynamical susceptibility $\chi_f$, which is exact in
$O(1/N^2)$.

The quality of the approximation  has been checked by comparison of
the static magnetic susceptibility $\chi_f(T)\equiv
\lim_{\nu\rightarrow 0} \chi(\nu)$ to  the {\em exact} Bethe-Ansatz
results\cite{Rajan83}, Fig.~\ref{fig-pnca-rajan-sus}. The \PNCAU\
susceptibility shows  a surpriselingly good agreement with the exact
results, even though the Bethe-Ansatz susceptibility  has been
obtained for the Coqblin-Schrieffer model. In particular, the
characteristic maximum for degeneracies $N\ge 3$  has been reproduced
in \PNCAU. On the other hand in {\em NCA} not even a saturation of $\chi_f$
occurs for $0.1T_K <T<T_K$ and $N=2$. The deficiencies of the {\em NCA} do
not show up so strongly at e.g.~$N=6$ or at higher temperatures
$T>T_K$, where vertex corrections become less important.

The scattering function $S(\nu)$ for  neutron scattering experiments
is linked to the magnetic susceptibility by the
dissipation-fluctuation theorem:
\begin{equation}\label{equ-s-fuc}
  S(\omega) \sim \frac{1}{1-\exp(-\beta\omega)}\Im m
\chi(\omega)\;\;\; .
\end{equation}
The right hand side approaches  the spin-fluctuation spectrum
$\sigma(\omega)\equiv \chi(\omega)/\omega$ for small $\omega$. In
Fig.~\ref{fig-sus-spinon}
%%(DISS 4.22)
 the spin-fluctuation spectra of the \PNCAU\  and the {\em NCA} are compared
for
three different degeneracies $N$ at a fixed temperature  $T=0.2 T_K$.
The pronounced maxima of $\sigma^{PNCA}(\omega)$ for $N>3$ resemble
the maxima in the static susceptibility discussed before. They
appear on the same energy scale  $\omega (T) \approx 0.5T_K$.
It is also interesting to note that even though no maximum is found
for $N=2$ a rather broad inelastic peak can be reported in
$\sigma^{PNCA}(\omega)$ which is not seen in the {\em NCA} spectrum.
Of course, this features disappears  well above $T_K$
for both approximations leaving a single elastic peak at $\omega=0$ in
the spectra.
The low frequency behaviour of $\Im m \chi(\omega)$ can be
interpolated by $\omega/(\Gamma_{neut}^2 + \omega^2)$.
$\Gamma_{neut}$ is the neutron scattering linewidth which is
experimentally defined by the position of the maximum. Here, it is
convenient  to use
\begin{equation}\label{equ-gamma-inter}
  \frac{1}{\Gamma^2_{neut}} = \lim_{\omega\rightarrow 0}\frac{\Im m
\chi(\omega)}{\omega} \;\; .
\end{equation}
In Fig.~\ref{fig-neutron}, the reduced linewidth $\Gamma_{neut}/T_K$
is displayed versus temperature for $N=2,4,6$ in \PNCAU. The linewidth
is nearly temperature independent in the local Fermi-liquid regime at
$T<T_K$. It develops a weak minimum  at $T\approx 0.5 T_K$ which is
slightly enhanced by increasing degeneracy and behaves like
$\sqrt{T}$ for high temperatures.

\section{Critical examination of the \PNCAU}

In the last three sections we discussed  \PNCAU\
results obtained for the Kondo regime of the model. In this section
we will focus our attention on the case $N=1$ and the mixed-valance
regime. For $N=1$ all diagrams are of the same order in the sense of
an $1/N$-expansion scheme. Nevertheless, this classification is
questionable here, since there is {\em no} Kondo effect left although
the diagram topology remains unchanged.
The one-particle Green's function is exactly known:
\begin{equation}\label{equ-res-level}
  F(z) = \frac{1}{z-\epsilon_f - \frac{V^2}{\# k}\sum_k
\frac{1}{z-\epsilon_k}} \;\; ;
\end{equation}
($\# k$ denotes the number of band electron states).
In figure \ref{fig-res-level-e1}
the one-particle spectra of the {\em NCA} and the \PNCAU\ are compared with
the exact result for two values of $\epsilon_f$.
While the $P_M(z)$ have been determined very accurately, oscillations
of the defect propagators $\xi(\omega) = \exp(-\beta\omega)
\frac{1}{\pi Z_f} \Im m P_M(\omega)$ during the iteration procedure
cause oscillations in the \PNCAU-spectra around the exact solution,
i.e.~convergence is not fully obtained in this case. These
oscillations contribute about 3\%  to the residual deviations of the \PNCAU.
In the most critical case, $\epsilon_f = -\Delta$, the  {\em NCA}-pathology is
strongly pathological near $\mu$, but the \PNCAU\ is much less so.
For $\epsilon_f = -3\Delta$ only a very weak pathology at $\mu=0$
remains, while the {\em NCA} still produces an unphysical resonance at the
chemical potential.

A rather strange two peak structure  is found in the spectrum for the
case of the intermediate valence regime $\epsilon_f = -\Delta$ and
$N=2$, shown in Fig.~\ref{fig-iv-e1}a:
the first peak nearly at $\mu = 0$ exceeds clearly the density of states
rule (\ref{equ-dos-rule}) $\rho(0) = 0.21$ for $n_f\approx 0.6$, which
would be the height of the minimum in between both peaks. We suppose
that the {\em true} spectrum will monotonically decrease from the
minimum position onwards for decreasing energy. The first peak clearly
reflects a \PNCAU-pathology already seen in Fig.~\ref{fig-res-level-e1}a.
Despite this spurious  structure,  the overall violation of the the
DOS-sum rule is  reduced from almost $60\%$ in {\em NCA} down to $15\%$ in
\PNCAU.
In part (b) of the figure, $\epsilon_f$ is chosen to be $+\Delta$.
Again, only a weak pathology is found in this case, and the spectrum
can be reasonably  fitted using a Lorentzian with a width of
$0.95\Delta$. This resembles the fact that there is no blocking
effect in this regime, whereas the local correlations still
reduce the spectral weight to  $<X_{00}> +
<X_{mm}> = 0.87<1$.

\section{Conclusion and Outlook}

We have demonstrated  that our new \PNCAU-approximation to the
Anderson-model improves the low energy properties in the Kondo-regime
quite remarkably, whereas the satisfactory behaviour of the old
{\em NCA}-approximation at higher energies is maintained. Vertex corrections
do have in fact a large impact on the low energy excitations:
position and height of the ASR for $N=2$ is shifted towards the
chemical potential as consistent  with Friedel's sum rule. Also the
magnetic susceptibility agrees surprisingly well with the Bethe-Ansatz
results. We plan to apply this method also to the finite-$U$
Anderson-model and the extended Anderson model
\cite{FreytagKel91,AndersQinGre92}, which includes a direct exchange
interaction between $f$- and conduction electrons. The self-consistent
set of equations have been derived already \cite{Anders95}.
Our perturbational approach to the Anderson-model
has opened the prospect of  studying the low temperature properties of Heavy
Fermion systems in much more detail  before. Quite generally, the
so-called LNCA scheme\cite{GrewePruKei88} can be applied, which provides
a very successful perturbational treatment of the Anderson
lattice. Within this theory based on a picture of independent
effective sides plus quasi-particle interactions collective effects
like super-conducting or magnetic ground states can be calculated in a
systematic way. In order to underline this perspective  we present the
temperature dependent pseudo-gap formation in the
$f$-spectrum of the Anderson lattice with $N=2$ and $\epsilon_f = -3$
in Fig.~\ref{fig-n2-l-pnca}. The temperatures are given in units of the
corresponding  impurity Kondo temperature which deviates moderately from
 the  characteristic temperature of the lattice $T^\ast$. While the
LNCA-scheme combined with the local {\em NCA} is limited to temperatures
larger than $T_K$ due to violations of  the local Fermi-liquid
properties, the so-called L-\PNCAU, which uses the \PNCAU\ to solve the
effective site, reaches much lower temperatures. Since now the ASR of
the impurity is located only very slightly above
the chemical potential in accordance with Friedel's sum rule, the
pseudo-gap is formed around $\omega =0$ in the L-\PNCAU. In a
forthcoming publication we will investigate the temperature dependent
quasi-particle band-structure and the impact on the transport
coefficients and optical conductivity in greater detail.

{\hfill *** \hfill}

The author is very grateful to Prof.~N.~Grewe for many fruitful
discussions on the subject of this work and careful reading of the
manuscript. He would also like to acknowledge valuable conversations
with H Keiter, J Keller, Th Pruschke and P W\"olfle.

\begin{table}[t]
\begin{center}
{\bf Friedel's Sum Rule and the  {\em NCA}-Theory:}

\vspace*{5truemm}
{\tabcolsep=5truemm
\begin{tabular}{c|ccc}
\hline
$T/T_K$ & $n_f$ & $ \dps  \sum_m
\frac{\delta_m}{\pi} $ & $\kappa_{res}$ \\
\hline\\
0.1 &      .8682     &       .5863     &       .2819      \\
0.2 &      .8682     &       .6114     &       .2567      \\
0.3 &      .8682     &       .6379     &       .2303       \\
\\
\hline
\end{tabular}
}
\end{center}

{\hfill (a)}

\begin{center}
{\bf Friedel's Sum Rule and the  \PNCAU-Theory:}

\vspace*{5truemm}
{\tabcolsep=5truemm
\begin{tabular}{c|ccc}
\hline
$T/T_K$ & $n_f$ & $ \dps  \sum_m
\frac{\delta_m}{\pi} $ & $\kappa_{res}$ \\
\hline\\
0.1 &      .8445     &       .9023     &      -.0577 \\
0.2 &      .8525     &       .9140     &      -.0615 \\
0.3  &      .8565     &       .9207     &      -.0641 \\
\\
\hline
\end{tabular}
}
\end{center}

{\hfill (b)}

\caption{Parameter:
$N=2,\epsilon_f=-3\Delta,W=10$}
\label{tab-friedel-pnca}
\end{table}

\newpage
\begin{figure}[ht]
\caption{Diagrammatic contributions to the self-energy $\Sigma_0(z)$ of
the unoccupied propagator $P_0$ up to  $V^8$.}
\label{fig-sig0}
\end{figure}

\begin{figure}[htb]
  \caption{Diagrammatical representation of the self-consistency
equation of the  vertex function $\Delta_m(x,y)$ in $O(1/N^2)$.}
\label{fig-vertex-voll}
\end{figure}

\begin{figure}[htb]
\caption{Generating functional for the self-energies and the Green's
functions up to order $O(1/N^2)$. The \PNCAU\ contains additional
higher order contributions.}
\label{fig-gen-func}
\end{figure}

\begin{figure}[htb]
  \caption{Comparison of the spectra of $P_0(\omega)$ (a) and
$P_m(\omega)$ (b) in NCA and  \PNCAU\ at $T=0.5, 1.0, 2.0 T_K$. The
increase in height at $\omega=0$ is correlated with a decrease of
temperature. Parameters: $N=2, \epsilon_f = -3\Delta, W=10\Delta$.}
\label{fig-prop-0-m}
\end{figure}

\begin{figure}[htb]
  \caption{Threshold behaviour of the spectrum of $P_0(z)$ (a) and an
estimation for the threshold exponent $\alpha_m$ (b). Parameters as before.}
\label{fig-schwell-a0}
\end{figure}

\begin{figure}[tb]
\caption{Specific heat contribution of the impurity vs temperature
$T/T_K$ for
(a) different values of the effective Anderson width
$\Delta^\star = \pi |V^\star|^2\rho_0$, $\epsilon_f=-3\Delta_0$, and (b)
different values of $\epsilon_f$. Parameters: $N=2,
\epsilon_f =-3\Delta_0, \Delta_0 = 0.1W$.}
\label{fig-sh-nca-n2-tk}
\end{figure}

\begin{figure}[tb]
\caption{Specific heat contribution of the impurity vs temperature
$T/\Delta_0$ for
(a) different values of the effective Anderson width
$\Delta^\star = \pi |V^\star|^2\rho_0$, $\epsilon_f=-3\Delta_0$, and (b)
different values of $\epsilon_f$. Parameters: $N=2,
\epsilon_f =-3\Delta_0, \Delta_0 = 0.1W$.}
\label{fig-sh-nca-n2-delta0}
\end{figure}

\begin{figure}[tb]
\caption{Temperature dependency of the ASR in (a) \PNCAU\ and  (b) in
NCA. Parameters: as before.}
\label{fig-asr-n2}
\end{figure}

\begin{figure}[ht]
\caption{Impact of the vertex correction on the one particle spectrum.
Parameters: as before.}
\label{fig-asr-vertex}
\end{figure}

\begin{figure}[ht]
\caption{Minimum of the self-energy  versus temperature for $N=2,4,6$.
Parameters: $ \epsilon_f(N=2) = -3\Delta,  \epsilon_f(N=4) = -5\Delta,
\epsilon_f(N=6) =
-6.7\Delta, W =10\Delta$.}
\label{fig-im-self}
\end{figure}

\begin{figure}[tb]
\caption{Diagrammatic representation of the self-consistency
condition of the magnetic vertex function $\Gamma(x,y)$.}
\label{fig-mag-sus-vertex}
\end{figure}

\begin{figure}[tb]
\caption{Static magnetic susceptibility calculated (a) in  NCA, (b) in
\PNCAU\ for the single impurity Anderson model and (c) for the
Coqblin-Schrieffer models using the Bethe Ansatz. The NCA results are
normalized with respect to the \PNCAU\ values at $T=0$.
Parameters: as in Fig.~10}
\label{fig-pnca-rajan-sus}
\end{figure}

\begin{figure}[tb]
\caption{$\Im m \chi(\omega)/\omega$ vs~$\omega/T_K$ for $N=2,4,6$,
calculated at $T=0.2 T_K$ (a)in \PNCAU\ and (b) in NCA.
Parameters: as in Fig.~10}
\label{fig-sus-spinon}
\end{figure}

\begin{figure}[tb]
\caption{Neutron scattering linewidth $\Gamma_{neut}$ vs.~$T/T_K$
calculated in \PNCAU\ for different $N$.
Parameters: as in Fig.~10}
\label{fig-neutron}
\end{figure}

\begin{figure}[tb]
\caption{One-particle spectra in NCA and \PNCAU\ compared with the {\em
exact solution} for (a) $\epsilon_f = -\Delta,  T= 1/15 \Delta$ and
(b) $\epsilon_f = -3\Delta,  T= 1/60 \Delta$. Parameter: $N=1$.}
\label{fig-res-level-e1}
\end{figure}

\begin{figure}[htb]
\caption{One-particle spectra in NCA und \PNCAU\  for (a) $\epsilon_f =
-\Delta$ and (b) $\epsilon_f = +\Delta$, the intermediate
valence regime. Parameters: $N=2, T= \frac{1}{15} \Delta$.}
\label{fig-iv-e1}
\end{figure}

\begin{figure}[tb]
\caption{$f$-spectra of the periodic  Anderson model in the vicinity
of the chemical potential (a) in L-\PNCAU\ and (b) in LNCA for three
temperature  $T=0.75, 1.0, 2.0 T_K$. Parameters $N=2,\epsilon_f=-3\Delta,
W=10\Delta$.}
\label{fig-n2-l-pnca}
\end{figure}

\end{document}